\documentclass[reprint,superscriptaddress,
nofootinbib,
amsmath,amssymb,
aps,showkeys,showpacs,
prd,
]{revtex4-2}

\usepackage{caption}
\usepackage{csquotes}
\usepackage{orcidlink}
\usepackage{footmisc}
\usepackage{subcaption}
\usepackage{graphicx}
\usepackage{dcolumn}
\usepackage{bm}
\usepackage{amsmath}
\DeclareMathOperator{\arctanh}{Arctanh}
\usepackage[T1]{fontenc} 
\usepackage{hyperref}
\hypersetup{
	colorlinks=true,
	linkcolor=blue,
	citecolor=red}

\begin{document}
	
	\title{$\Lambda$vCF: Extending $\Lambda$CDM into a unified model with particle creation}
	
	\author{Vishnu A Pai \orcidlink{0000-0003-4161-3383}
	}
	\email{vishnuajithj@gmail.com}
	\affiliation{Department of Physics, Cochin University of Science and Technology, Cochin -682022}
	
	\author{Titus K Mathew \orcidlink{0000-0002-9307-3772}
	}
	\email{titus@cusat.ac.in}
	\affiliation{Department of Physics, Cochin University of Science and Technology, Cochin -682022}
	
	\date{\today}

\begin{abstract}
	We present a novel extended version of the $\Lambda$CDM model that provides analytical solution for Hubble parameter uniting all epochs of cosmic evolution starting from inflation to late-acceleration, with intermediate radiation and matter-dominated epochs. This is achieved by relaxing the perfect fluid assumption in the standard model and considering a general viscous cosmic fluid (vCF) with non-zero particle creation rate and evolving adiabatic equation of state. Transition points of the Universe and the finite boundary connecting them is exactly determined. We then propose a novel method to determine the early-time viscous coefficient and inflation energy scale using the Cosmic Mode Index value postulated by Padmanabhan. Considering the data from the Planck 2018 analysis, this yields an inflationary Hubble parameter of $H_{I}\approx10^{13}$\,GeV. An equivalent scalar-field description for the inflationary epoch is then constructed and inferences are made regarding the nature of inflation. Notably, we find that the model describes an ultra-slow-roll hilltop inflation scenario with a graceful exit to radiation-dominated epoch. Subsequently, we show that bulk viscosity in this model can be expressed as Israel-Stewart equation in relativistic dissipative hydrodynamics with an appropriate underlying viscous coefficient and relaxation time that satisfy the causality constraint in its extreme limit. Finally, by comparing the evolution of this causal relation and its Navier-Stokes counterpart, we infer that the evolution from inflation to radiation era signifies a fluid transitioning from viscoelastic to pseudoplastic behavior.
\end{abstract}

\maketitle
\section{Introduction}

Deriving a unified cosmological model, which describes the entire evolution of the universe right from the early inflation to the late acceleration through the radiation and subsequent matter dominated epochs is important, as it enables one to better understand the entire expansion history of the universe with its smooth transition from one epoch to another. In addition to providing better constraints on cosmological observables, such a model can also uncover additional insights about the curious connection between the early and late epochs of the universe. The major challenge in obtaining such a cosmological model is producing a viable early inflationary epoch which smoothly transitions into the subsequent radiation era followed by matter epoch and finally, the late accelerated epoch. In the recent literature, there have been attempts to create such unified cosmological models; however, to obtain analytical solutions for the Hubble parameter \cite{Moreno-Pulido:2022upl,doi:10}, these studies consider an arbitrary fixing of the equation-of-state parameter of the effective energy density during different epochs, instead of letting it evolve smoothly with cosmic history. This is done so that the cosmic fluid mimics the behavior of dominant cosmic component (i.e. one considers $c_s^2=w=1/3$ to study infaltion-radiation regime, and then $w=0$ to study late time matter-dark energy era). This means, such models actually fail to predict a smooth transition connecting all the major dominant epochs of the universe.

In this article, we propose a suitable extension of the standard cosmological model that can predict a smooth evolution of the universe that unifies all major epochs of cosmic evolution, i.e. a continuous transition from early inflation, to radiation, to matter domination and finally to late acceleration epoch. Since the standard $\Lambda$CDM model agrees well with observations, we design our model by including the cosmological constant, so that the late-time dynamics matches closely with the standard $\Lambda$CDM. The main challenge is then to suitably model the early inflationary phase and determine a equation of state that ensures smooth transitions into later stages of cosmic evolution. To drive the early inflationary expansion, we consider the mechanism of bulk viscosity emerging from an isentropic particle creation process, and to model a smooth transition from radiation to matter phase, we adopt a simple, evolving barotropic equation of state for the effective cosmic fluid in a suitable way. The use of bulk viscosity to drive early inflation has the added advantage that reheating can occur naturally as a result of the viscous effects. In this sense, our model is the most straight forward judicious extension of the $\Lambda$CDM model that incorporates bulk viscosity. This extended version of the standard $\Lambda$CDM model that  incorporates viscosity arising from particle creation is referred to as the $\Lambda$vCF model ($\Lambda$–viscous Cosmic Fluid model).

It is well established that bulk viscosity arising in the cosmic fluid can generate the negative pressure that is required for driving the accelerated expansion of the universe \cite{Gron:1990ew,PADMANABHAN1987433}. This interesting dynamical behavior arises in scenarios where the fluid is far-from-equilibrium and its relaxation time exceeds the Hubble time, which is the characteristic expansion timescale. In the conventional dissipative bulk viscous models, it is difficult to justify a relaxation time (which represents the microscopic interaction timescale between fluid constituents) that is larger than the Hubble time. Therefore, in the present work, we attribute the origin of the viscous pressure to an isentropic particle creation process, which has been hypothesized to have occurred in the early universe \cite{Parker:1971pt,PhysRevD.30.2036,Nojiri:2003vk}. Within this interpretation, having a relaxation time that is longer than the Hubble time is physically justified, as this is a non-dissipative process and the relaxation time in this case corresponds to the finite time duration of the particle creation process \cite{PhysRevD.64.063501}. To model the generation of bulk viscosity through particle creation in the early universe, we adopt the thermodynamic approach developed in \cite{mnras/266.4.872,PhysRevD.54.6101,PhysRevD.64.063501}, where particle creation is formulated within the context of relativistic dissipative hydrodynamics. Notably, in \cite{PhysRevD.64.063501} author further demonstrates that this formalism is consistent with kinetic theory approach.

The article is organized as follows: In Sec. \ref{1} we will model the evolution of flat FLRW Universe in the presence of bulk viscous fluid with non-zero particle creation. In Sec \ref{2} we study the influence of bulk viscosity in the evolution of some relevant cosmological observables such as transition redshift,  number of perturbation modes crossing the Hubble horizon and determine the exact boundary connecting the two asymptotic de Sitter epochs. In Sec \ref{3} we construct an equivalent scalar field description for the inflationary expansion, determine the evolution of slow-roll parameters, nature of inflation and assess the possibility of graceful exit. Sec \ref{isp} is dedicated to investigating the possibility of having an equivalent causal evolution equation for bulk viscosity, and determining the dissipative nature of viscous fluid. Finally, in Sec \ref{5} we summarize the results of the entire article.
	
\section{$\Lambda \text{v}$CF model of the Universe}\label{1}

In a spatially flat, homogeneous and isotropic space-time, the Friedmann equations describing the evolution of the Universe in the presence of a bulk viscous cosmic fluid (vCF) with energy density `$\rho$', local equilibrium pressure `$p$', bulk viscous pressure `$\Pi$', and the cosmological constant `$\Lambda$' takes the form,
\begin{align}
	3H^2=&\;\rho+\Lambda\label{F1}\\
	2\dot{H}+3H^2=&\;-p -\Pi +\Lambda \label{F2}
\end{align}
Here, `$a$' is the scale factor, and $H=\dot{a}/a$ is the Hubble parameter of the Universe, and we have set, $8\pi G/c^2=1$. Also, over-dot represents the derivative with respect to cosmic time `$t$'. For a fluid with non-zero particle creation the continuity equations are given as \cite{PhysRevD.61.083511},
\begin{eqnarray}
	&\dot{n} +3Hn =\Gamma n\label{consn}\\
	&\dot{\rho}+3H\left(\rho+p+\Pi\right)=0\label{cons}
\end{eqnarray}
Here, $n$ is the particle number density and $\Gamma$ represents the particle creation rate. Interestingly, at the minimal level, the particle creation process can be isentropic, such that particles of constant entropy are created. As a result the particle creation rate and bulk viscous pressure can be linearly related \cite{PhysRevD.64.063501}. This can be seen by combining the conservation equations given above, with Gibbs equation in causal thermodynamics,
\begin{equation}\label{Tds}
	Tds=d\left(\rho/n\right) + pd\left(1/n\right)
\end{equation}
with `$s$' being the specific entropy. Combining equations (\ref{consn}), (\ref{cons}) and (\ref{Tds}), we get, 
\begin{equation}\label{Gamma}
	\Gamma=-\frac{3\Pi H}{\rho + p}
\end{equation}
Hence, if evolution of viscous pressure is known, then one can directly associate it with a particle creation rate. Also, $\Pi<0 \implies \Gamma>0$ which means particles are being created, while $\Pi>0 \implies \Gamma<0$, signifying particle number reduction. Hence, one can either model particle creation rate and determine the evolution of $\Pi$, or conversely, specify the evolution of $\Pi$ and then determine the corresponding particle creation rate. In the present study we consider the latter approach. Subsequently, the set of equations governing the evolution of the Universe becomes complete once the evolution of the bulk viscous pressure and the equation of state for vCF are specified, as done below.

In the early Universe, the effective energy density contains contributions from matter (both baryonic and dark), radiation, as well as their mutual interactions. However, since exact treatment of each component along with their possible mutual interaction is complicated, we account the collective evolution of the fluid by assuming an effective barotropic equation of state that varies with expansion scale of the Universe. One such equation of state often considered in literature is\footnote{Special case of this equation of state can be directly obtained by considering matter-radiation mixture, $\omega=(p_m+p_r)/(\rho_m+\rho_r)$. And if, $p_m=0$ and $p_r=\rho_r/3$, we get $\omega=(1/3)/[1+(\rho_m/\rho_r)a]$} \cite{deLeon2012},
\begin{equation}\label{eos}
	\omega= \frac{1}{3}\left[\frac{1}{1+ \Omega^{0}_{\gamma} a^{\gamma}}\right]
\end{equation}
Here, $\gamma\geq0$ hereafter called the swiftness parameter, indicates the pace with which vCF transitions from relativistic to non-relativistic regime and, $\Omega^{0}_{\gamma}=\Omega^{0}_{m}/\Omega^0_r$, is the ratio of critical energy densities of non-relativistic and relativistic matter component. This equation of state is often considered in literature for explaining a transition from radiation to matter dominated era.

Evolution of $\Pi$ is then determined from the relation, $\Pi=-\zeta\nabla_\mu u^{\mu}$, where $u^{\mu}$ is the four-velocity of the fluid in the comoving frame and $\zeta$ represents the bulk viscous coefficient \cite{PhysRev.58.919}. Even though this expression follows from Eckart's theory (which is acausal), in Sec. \ref{isp} we will show that this viscous pressure can identically emerge as the solution of an Israel-Stewart type equation having a particular relaxation time and redefined bulk viscous coefficient, such that causality is respected. To describe the evolution of the universe, one must then postulate the form of viscous coefficient. In literature, one often models coefficient of bulk viscosity to be: (a) power law function in energy density of the fluid \cite{PhysRevD_53}, (b) function of expansion rate of the Universe \cite{Ren:2005nw}. Nonetheless, one can consider a more general case where, particle creation processes inherently depend on both energy density of the cosmic fluid and the expansion velocity (i.e., Hubble velocity). In which case, it is obvious to consider the viscous coefficient to be explicitly proportional to both $\rho$ and $H$. With this very basic postulate, one arrives at;
\begin{equation}\label{zeta}
	\zeta \propto \rho \;\;\; \& \; \;\;\zeta \propto H \implies \zeta=\zeta_{0}\rho H/H^2_p
\end{equation}
Here, $\zeta_{0}$ is a constant dimensionless parameter and $H^2_p$ is inverse Planck time square. In the above relation, $H^2_p$ is included for dimensional consistency, and since we are considering viscous pressure arising in the early Universe --when size is close to, but still above, the Planck scale-- it is natural to adopt the Planck time as the dimensional constant in Eqn. (\ref{zeta}), which enables us to express the value of $\Pi$ relative to the Planck energy scale.

Combining the Friedmann equations (\ref{F1}) and (\ref{F2}) with the above ansatz, i.e. Eqns. (\ref{eos}) and (\ref{zeta}), we obtain the non-linear ordinary differential equation (ODE),
\begin{equation}\label{ODE}
	2\dot{H}=\left[3H^2-\Lambda \right]\left\{3\zeta_0\frac{H^2}{H^2_p}-\left[1+\frac{1/3}{1+\Omega^0_{\gamma} a^{\gamma}}\right]\right\}
\end{equation}
We can then rewrite the above equation by setting scale factor as the variable by using relation $d/dt = aHd/da$. Subsequently, we can then divide both sides by $3H_{0}^2$, where $H_0$ is the present value of $H$, to arrive at,
\begin{equation}\label{ODE2}
	\hspace{-0.14cm}\frac{2a\mathcal{H}}{3}\frac{d\mathcal{H}}{da}=\left[\mathcal{H}^2-\Omega^{0}_{\Lambda} \right]\left\{\Omega^{0}_{\zeta}\mathcal{H}^2-\left[1+\frac{1/3}{1+\Omega^0_{\gamma} a^{\gamma}}\right]\right\}
\end{equation}
with dimensionless parameters, $\mathcal{H}=H/H_{0}$ and,
\begin{equation}
	\Omega^0_{\Lambda}=\frac{\Lambda}{3H_0^2}\;\;\;\textbf{;}\;\;\Omega^0_{\zeta}=3\zeta_{0}\left[\frac{H_0}{H_p}\right]^2\;\; \textbf{;}\;\; \Omega^0_{\gamma}=\frac{\Omega^0_{m}}{\Omega^0_{r}} \label{red}
\end{equation}
Interestingly, in spite of being a non-linear ODE, Eqn. (\ref{ODE2}) provides an analytical solution to Hubble parameter, which, under the condition $\mathcal{H} \to 1$ as $a \to 1$, becomes,
\begin{equation}\label{Hsolg}
	\mathbf{\mathcal{H}=\sqrt{\frac{\Omega^0_{\Lambda} +\left(1-\Omega^0_{\Lambda}\right) \left\{\Upsilon_i\, a^{-\Theta} + \Upsilon_j\, a^{-\left(\Theta+\gamma\right) } \right\}}{1 + 3\Omega^0_{\zeta}\left(1-\Omega^0_{\Lambda}\right) \Upsilon_k \,a^{-\left(\Theta+\gamma\right) }}}}
\end{equation}
where, we have defined,
\begin{align}
    \Theta=&\,4-\gamma-3\Omega^{0}_{\zeta} \Omega^{0}_{\Lambda} \label{Delta}\\
	\Psi=&\,\sqrt[\gamma ]{\Omega^{0}_{\gamma} +1} - 3 \Omega^{0}_{\zeta} (1-\Omega^{0}_{\Lambda} )\left[\frac{\Omega^{0}_{\gamma} \mathfrak{F}_{C}}{\Theta} + \frac{\mathfrak{F}_{A}}{\Theta +\gamma }\right]\\
    \Upsilon_i=&\,\frac{1}{\Theta \Psi}\left(3\Omega^0_{\zeta}\Omega^0_{\gamma}\Omega^0_{\Lambda}\mathfrak{F}_{D}\right)\\
	\Upsilon_j=&\,\frac{1}{\Theta \Psi}\left[\sqrt[\gamma ]{\Omega^{0}_{\gamma}  a^{\gamma }+1} + \frac{3\Theta\Omega^0_{\zeta}\Omega^0_{\Lambda} \mathfrak{F}_{B}}{\Theta+\gamma}\right]\\
	\Upsilon_k=&\,\frac{1}{\Theta \Psi} \left[ \Omega^{0}_{\gamma}\mathfrak{F}_{D} + \frac{\Theta\mathfrak{F}_{B}}{\Theta+\gamma}\right]
\end{align}
\begin{align}
	\mathfrak{F}_A=&\,_2\mathcal{F}_1\left[\frac{\gamma -1}{\gamma },\frac{-\left(\Theta+\gamma\right)}{\gamma };\frac{-\Theta}{\gamma };-\Omega^0_{\gamma} \right] \label{FA}\\
	\mathfrak{F}_B=&\,_2\mathcal{F}_1\left[\frac{\gamma -1}{\gamma },\frac{-\left(\Theta+\gamma\right)}{\gamma };\frac{-\Theta}{\gamma };-\Omega^0_{\gamma}  a^{\gamma }\right] \label{FB}\\
	\mathfrak{F}_C=&\,_2\mathcal{F}_1\left[\frac{\gamma -1}{\gamma },\frac{-\Theta}{\gamma };\frac{\gamma-\Theta}{\gamma };-\Omega^0_{\gamma}\right] \label{FC}\\
	\mathfrak{F}_D=&\,_2\mathcal{F}_1\left[\frac{\gamma -1}{\gamma },\frac{-\Theta}{\gamma };\frac{\gamma-\Theta}{\gamma };-\Omega^0_{\gamma} a^{\gamma }\right] \label{FD}
\end{align}
Here, $_2\mathcal{F}_1[x_i]$ denotes the \enquote{\textit{hyper-geometric}} function. Even-though this solution might seem complicated and lengthy, one must note that having analytical solution to an ODE is always preferred as it enables one to study the exact behavior of the system without having to deal with truncation and round-off errors that numerical solutions carry. Also, for a given constant value of $\gamma$, this reduces to a simple form. For instance, in the case $\gamma=1$, we get,
\begin{equation}\label{Hsol}
	\mathcal{H}=\sqrt{\frac{\Omega ^0_{\Lambda }+\bar{\Omega}^0_r \,a^{3 \Omega ^0_{\zeta } \Omega ^0_{\Lambda }-4}+\bar{\Omega}^0_m a^{3 \Omega ^0_{\zeta } \Omega ^0_{\Lambda }-3}}{1+\Omega ^0_{\zeta } \left[\bar{\Omega}^0_r \, a^{3 \Omega ^0_{\zeta } \Omega ^0_{\Lambda }-4}+\bar{\Omega}^0_m \, a^{3 \Omega ^0_{\zeta } \Omega ^0_{\Lambda }-3}\right]}}
\end{equation}
with redefined model parameters,
\begin{align}
	\mathcal{X}=&\,1+\Omega^0_{\gamma}-\Omega^0_{\gamma } \Omega^0_{\zeta }\\
	\bar{\Omega}^0_m=&\,\frac{\Omega^0_{\gamma} \left(1-\Omega^0_{\Lambda }\right) \left(4-3 \Omega^0_{\zeta } \Omega^0_{\Lambda }\right)}{4\mathcal{X}-\Omega^0_{\zeta } \Omega^0_{\Lambda } \left(1+3\mathcal{X}-3 \Omega ^0_{\zeta }\right)-3 \Omega^0_{\zeta }} \label{Ome}\\
	\bar{\Omega}^0_r=&\,\frac{4\left(1-\Omega^0_{\Lambda }\right) \left(1-\Omega^0_{\zeta } \Omega^0_{\Lambda }\right)}{4\mathcal{X}-\Omega^0_{\zeta } \Omega^0_{\Lambda } \left(1+3\mathcal{X}-3 \Omega^0_{\zeta }\right)-3 \Omega^0_{\zeta }}  \label{Ore}
\end{align}
Note that in the limit $\Omega^0_{\zeta} \to 0$, both the general solution and the special case, becomes similar to $\Lambda$CDM model in the late-phase of the Universe. However, it differs slightly due to the presence of the model parameter $\Omega^0_{\gamma}$ which arises from effective equation of state assumed in Eqn. (\ref{eos}). From Eqns. (\ref{Ome}) and (\ref{Ore}), it is clear that, larger the value of $\Omega^0_{\gamma}$, more closer the model is to $\Lambda$CDM case. And from observations \cite{aghanim2020planck}, one obtains $\Omega^0_{\gamma} \approxeq 10^4$. Hence, the present model resembles the standard $\Lambda$CDM model in the late-phase. However, in the early epoch near matter-radiation equality, significant deviations from the standard model can arise depending on the value of swiftness parameter $\gamma$ and the critical energy density ratio $\Omega^0_{\gamma}$. More importantly, notice that according to both Eqn. (\ref{Hsolg}) \& (\ref{Hsol}), this model predicts a Universe evolving from an initial quasi-de-Sitter accelerated expansion driven by $\Pi$, followed by two decelerated expansion regimes driven by radiation, and matter component respectively, which then transitions into a late-accelerated expansion epoch driven by $\Lambda$.
\begin{figure*}
	\centering
	\begin{subfigure}{\columnwidth}
		\centering
		\includegraphics[width=\columnwidth]{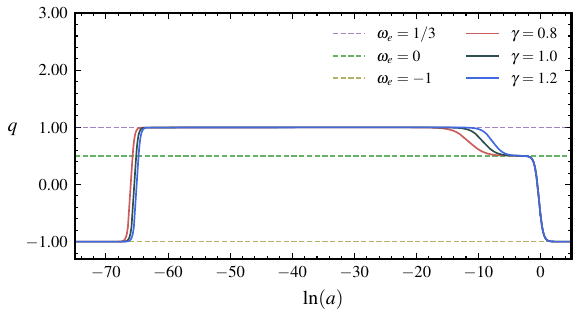}
		\caption{For different $\gamma$ with $\Omega^0_{\gamma}=10^4$.}
		\label{decsub1}
	\end{subfigure}
	\begin{subfigure}{\columnwidth}
		\centering
		\includegraphics[width=\columnwidth]{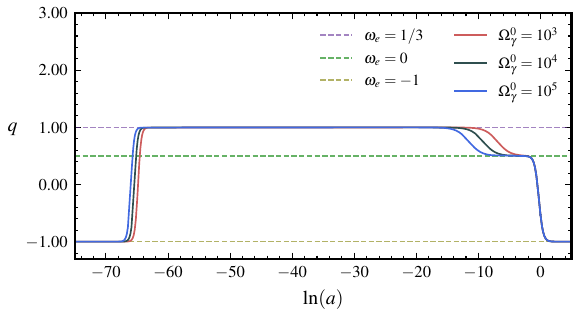}
		\caption{For different $\Omega^0_\gamma$ with $\gamma=1$.}
		\label{decsub2}
	\end{subfigure}
	\caption{Evolution of deceleration parameter with $x=\ln a$, for different values of model parameters. For plotting, we consider the values; $\Omega^0_{\Lambda}=0.68$ \cite{aghanim2020planck} and $\Omega^0_{\zeta}=10^{-108}$. Also, $\omega_e=-(1+2\dot{H}/3H^2)$ denotes effective equation of state.}
	\label{dec}
\end{figure*}

The two asymptotic de-Sitter solutions of this model can be obtained, either from the exact solution (\ref{Hsolg}), or by considering $d\mathcal{H}/da = 0$ in Eqn. (\ref{ODE2}) and applying the limits $a\to0$ or $a\to \infty$ respectively. Accordingly we get,
\begin{align}
	\mathcal{H}\,\big|_{a\to \infty}&=\mathcal{H}_{\Lambda}=\,\frac{H_{\Lambda}}{H_0} \approxeq \sqrt{\Omega^0_{\Lambda}}&\label{hl}\\
	\mathcal{H}\,\big|_{a\to 0}&=\mathcal{H}_{I}=\,\frac{H_I}{H_0}\approxeq\;{\sqrt{\frac{4}{3\Omega^0_{\zeta}}}}&\label{hi}
\end{align}
While the late-time de-Sitter solution (i.e. Eqn. (\ref{hl})) is identical to one obtained in $\Lambda$CDM model, the early de-Sitter solution (Eqn. (\ref{hi})) is obtained as a direct consequence of considering bulk viscosity given in Eqn. (\ref{zeta}), and it depends only on viscous coefficient $\Omega^0_{\zeta}$.  

\section{Analytical Determination of Cosmological Transition Points} \label{2}

Substituting Eqn. (\ref{ODE}) in the standard definition of deceleration parameter, $1+q=-\dot{H}/H^2$, we obtain the expression,
\begin{equation}\label{q}
	1+q=-\frac{3}{2}\left[1-\frac{\Omega^{0}_{\Lambda}}{\mathcal{H}^2} \right]\left[\Omega^{0}_{\zeta}\mathcal{H}^2-\left(1+\frac{1/3}{1+\Omega^0_{\gamma} a^{\gamma}}\right)\right]
\end{equation}
Using the obtained analytical solution (\ref{Hsolg}), we can plot the evolution of the deceleration parameter as shown in Fig. (\ref{dec}). For this, we consider the parameter values obtained from Planck analysis \cite{aghanim2020planck}, i.e. $\gamma=1$, $\Omega^0_{\Lambda}=0.68$ and $\Omega^0_{\gamma}=10^4$, and the value of viscous coefficient as $\Omega^0_{\zeta}\approxeq10^{-108}$. {\it{The value of the bulk viscous coefficient provided here is determined via a novel approach that makes use of the value of cosmic mode index proposed by Padmanabhan, and the details will be presented in the next section.}} Clearly, one can see that the model predicts a Universe with an initial quasi-de-Sitter expansion which smoothly transitions into radiation dominated deceleration phase at some early transition redshift $z_e$ which is later followed by a matter dominated epoch and transition into a late accelerated expansion at redshift $z_t$. Interestingly, we find that the value of model parameters associated with swiftness of transition from radiation to matter epoch, also has a significant impact on the value of $z_e$. That is, from Fig. (\ref{decsub1}) and (\ref{decsub2}) we see that;
\begin{itemize}
	\item For fixed value of $\Omega^0_\gamma$ \textbf{:} $z_e$ decreases with increase in $\gamma$. Which means universe transition from acceleration to deceleration at some later time for a larger value of $\gamma$, as compared to a smaller value. 
	\item  For fixed value of $\gamma$ \textbf{:} $z_e$ increases with increase in $\Omega^0_\gamma$. This means universe transition from acceleration to deceleration at some earlier time for a larger value of $\Omega^0_\gamma$, as compared to a smaller value. 
\end{itemize}
Hence, $\gamma$ and $z_e$ are negatively correlated, while $\Omega^0_\gamma$ and $z_e$ are positively correlated parameters. Also, it is worth noting the fact that, smaller values of $\Omega^0_{\zeta}$ leads to larger transition redshifts ($z_e$) in the early universe.

\paragraph*{\textbf{Transition points:}} To determine the exact redshift at which the Universe transitions from acceleration to deceleration in the early Universe, we can equate the Hubble parameter relations obtained by setting $q=0$ in Eqn (\ref{q}), with the exact solution (i.e., Eqn. (\ref{Hsolg})) evaluated at the transition point $a=a_e$. Considering, $q=0$ in Eqn. (\ref{q}) and neglecting $\Omega^0_{\Lambda}/\mathcal{H}^2$ term (since, $\Omega^0_{\Lambda}/\mathcal{H}^2\ll1$ in the early Universe.) we get,
\begin{equation}\label{q0}
	\mathcal{H}_e^2=\frac{1}{3\Omega^{0}_{\zeta}}\left[\frac{2+\Omega^0_{\gamma} a_e^{\gamma}}{1+\Omega^0_{\gamma} a_e^{\gamma}}\right] \xrightarrow[]{\text{\scriptsize as $a_e \to 0$}}	\mathcal{H}_e^2=\frac{2}{3\Omega^{0}_{\zeta}},
\end{equation}
where $\mathcal{H}_e$ is the value of the Hubble parameter at the early time transition point corresponds to the transition from inflation to radiation epoch. Now let us obtain the $\mathcal{H}_e,$ directly from the general solution for the Hubble parameter by taking $a=a_e.$ Here we will neglect the cosmological constant from the solution (\ref{Hsolg}) as it has no relevant role in the early epoch, and recalling that, $\Omega^0_{\zeta}\ll1\ll\Omega^0_{\gamma}$ we get,
\begin{equation}\label{Hsole}
	\mathcal{H}_e^2=\frac{4 \sqrt[\gamma]{\Omega^0_{\gamma}  a_e^{\gamma }+1}}{4 \,a_e^4\sqrt[\gamma]{\Omega^0_{\gamma}}+ \Omega^0_{\zeta}\left[\,a_e^4 \mathfrak{F}_E - \,\mathfrak{F}_F+4 \sqrt[\gamma]{\Omega^0_{\gamma}  a_e^{\gamma }+1}\,\right]  } \notag
\end{equation}
\begin{align}
	\text{with, \;\quad}&\mathfrak{F}_E=\,_2\mathcal{F}_1\left[-\frac{4}{\gamma },\frac{\gamma-1}{\gamma};\frac{\gamma-4}{\gamma };- \Omega^0_{\gamma} \right] \label{FE}\\
	&\mathfrak{F}_F=\,_2\mathcal{F}_1\left[-\frac{4}{\gamma },\frac{\gamma-1}{\gamma};\frac{\gamma-4}{\gamma };- a_e^{\gamma} \Omega^0_{\gamma} \right]  \label{FF}
\end{align}
Since, $a_e$ is extremely small, one gets the approximate relations;  $\mathfrak{F}_F\approx1$, $\Omega^0_{\gamma}  a_e^{\gamma }\ll1 $, $\Omega^0_{\gamma}\gg1$ which implies, $ 1+\Omega^0_{\gamma}\approx\Omega^0_{\gamma}\;$ \& $\;\Omega^0_{\gamma}  a_e^{\gamma }+1 \approx 1$. With these approximations, the above solution reduces to,
\begin{equation}\label{He}
	\mathcal{H}_e^2=\frac{4}{ 3\Omega^0_{\zeta} + 4 \,a_e^4 \, (\Omega^0_{\gamma})^{1/\gamma }}
\end{equation}
Then by equating the relations (\ref{q0}) \& (\ref{He}) at $a=a_e$, we obtain scale factor at the transition point as,
\begin{equation}
	a_e = \sqrt[4]{\frac{3 \Omega^0_{\zeta}/4}{\sqrt[\gamma]{\Omega^0_{\gamma}}}} \implies x_e\approx\frac{\ln(\Omega^0_{\zeta})}{4} - \frac{\ln(\Omega^0_{\gamma})}{4\gamma}-0.072 \label{ae}
\end{equation}
where, $x_e=\ln(a_e)$. Substituting for the parameters we can obtain the magnitude of $a_e.$ To verify that, this is in conformity with the plot Fig. (\ref{dec}), we can consider the values; $\gamma=1$, $\Omega^0_{\zeta}\approxeq10^{-108}$ and $\Omega^0_{\gamma}=10^4$, in the above relation, from which one gets $x_e\approx -65$. Which matches exactly with the value seen in Fig. (\ref{dec}).

Similarly, one can also determine the transition point in the late phase of the Universe by equating the value of $\mathcal{H}$ obtained from Eqn. (\ref{Hsolg}), by considering the late-phase approximation, at which the parameter $\Omega_{\zeta}^{0}$ is not relevant, with the value of $\mathcal{H}$ obtained by setting $q=0$ in Eqn. (\ref{q}). The scale factor at the late-time transition point ($a_t$) and the corresponding value of the Hubble parameter ($\mathcal{H}_e$) at that point are then obtained as;
\begin{equation}\label{late}
	a_{t}= \sqrt[3]{\frac{1-\Omega^0_{\Lambda}}{2\Omega^0_{\Lambda}}}\quad\quad\textbf{\&}\quad\quad \mathcal{H}^2_{t}=3\Omega^0_{\Lambda}
\end{equation}
Using $\Omega^0_{\Lambda}=0.68$ we get $\mathcal{H}_t=1.42$ and $a_t=0.61$, which corresponds to a transition redshift of $z_t=0.63$. 

\begin{table}
	\renewcommand{\arraystretch}{1.4}
	\centering
	\begin{tabular}{|c|c|c|}\hline
		Dominating &\;Scale factor\;& \;Cosmic time duration $(\mathcal{T})$\;\\
		Epoch & interval & \;in appropriate units\;\\ \hline
		Inflation &$a_i \to a_e$& $1.39 \times 10^{-36}$ \textbf{s}\\
		Radiation &$a_e \to a_{eq}$& $6.81 \times 10^{4}$ \textbf{yrs}\\
		Matter &$a_{eq} \to a_t$& $7.86 \times 10^{9}$ \textbf{yrs}\\
		Dark energy &$a_t \to a_0$& $6.24  \times 10^{9}$  \textbf{yrs}\\\hline
	\end{tabular}
    	\caption{\label{tab1}Cosmic time elapsed in each dominant epoch of evolution of the Universe. Total age of the Universe is approximately $14.1$ Gyrs, which is slightly higher than $\Lambda$CDM prediction, which is $\approx 13.79$ Gyrs \cite{aghanim2020planck}}
\end{table}

Furthermore, since the present model offers a smooth evolution connecting all major expansion phases of the universe, we can determine the time duration of each dominant cosmic epoch, and the total age of the Universe using the equation: $\mathcal{T}=\int_{a_1}^{a_2} da/(aH),$ where $a_1$ and $a_2$ represents the initial and the final scale factor values. Considering, $H_0=67.4$ Km/s/Mpc, $z_{eq}=3400$, $\gamma=1$, $\Omega^0_{\Lambda}=0.68$, $\Omega^0_{\zeta}=10^{-108}$ and $\Omega^0_{\gamma}=10^4$, we tabulated the time elapsed during each epoch in Table. (\ref{tab1}).

\section{Constraining the bulk viscous coefficient using Cosmic Mode Index}

One of the key parameters that governs the dynamics of the early universe in the present model is the viscous coefficient, $\Omega_{\zeta}^0$, given in Eq. (\ref{red}), which actually drives the inflationary expansion. From Eq. (\ref{hi}), it is clear that this parameter determines the energy scale at the onset of inflation. Traditionally, parameters related to the early universe are determined using CMB data. However, in this work, we suggest a new approach to constrain the bulk viscous coefficient (and thus the inflation energy scale) based on Padmanabhan’s postulate that the total number of perturbation modes crossing the Hubble horizon during the entire evolution of the universe remains constant and is equal to $4\pi$\cite{doi:10.1142/S0218271813420017}. This invariant number is called the Cosmic Mode Index or CosMIn. Since the present model predicts the evolution of the universe from the early inflationary phase to the late-accelerating phase, through successive radiation and matter-dominated epochs, it gives a better method to obtain the cosmic mode index, which in turn constraint the viscosity coefficient and also the inflation energy scale.

\begin{figure}
	\centering
	\includegraphics[width=\columnwidth]{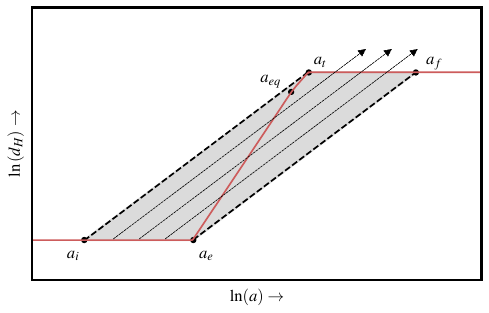}
	\caption{Evolution of logarithm of Hubble radius with $\ln(a)$. In this plot, arrowed lines represents the crossing of perturbation modes, and thick dashed black lines are tangent lines drawn at transition points of the Universe.}
	\label{NCF}
\end{figure}

During the early inflationary phase, the Hubble radius ($d_H=1/H$) remains constant, while the wavelengths of perturbation modes ($\lambda$), which are generated well inside the horizon, grows in proportion to the exponentially increasing scale factor of expansion. Once the wavelength of the perturbation mode exceeds the Hubble radius, it exits the Hubble horizon. Later, when the universe enters the decelerating phase, dominated by radiation or matter, the Hubble radius begins to grow, $d_H \propto a^2$ during the radiation-dominated era and $d_H \propto a^{3/2}$ during the matter-dominated era, allowing the previously exited modes to re-enter the Hubble volume. In the absence of a late de-Sitter phase, every mode which exited the Hubble volume during inflation, would re-enter in due time. However, the presence of the late de-Sitter epoch limits the number of modes re-entering the Hubble radius. As the universe transit into the late accelerated epoch, the Hubble horizon once again becomes stationary, as a result the modes that are re-entered the Hubble horizon during radiation/matter dominated epochs will exit the horizon during the late de-Sitter epoch. It has been shown that\cite{doi:10.1142/S0218271813420017}, the number of modes that exited the Hubble radius during inflation, the modes that re-entered the Hubble sphere during the radiation-matter era and the number of modes that will exit the Hubble horizon during the late de-Sitter phase are all equal to each other and is thus an invariant quantity. What is even more interesting is that, this constancy in the number of modes crossing the horizon, can be used to relate the inflation energy scale to the cosmological constant which drives the late accelerated phase of the Universe. A geometrical picture of the modes that cross the Hubble sphere during the evolution of the Universe can be obtained from Fig. (\ref{NCF}). In the figure, we have denoted the value of `$a(t)$' at the beginning of inflation as $`a_i$', end of inflation as $`a_e$', start of late-accelerated expansion as $`a_{t}$', and the boundaries of the late de-Sitter phase as $`a_f$'. We will denote the Hubble radius during these respective scale factor values as, $d_{\mathcal{H}_{i}}$, $d_{\mathcal{H}_{e}}$, $d_{\mathcal{H}_{t}}$ and $d_{\mathcal{H}_{f}}$ respectively.  Boundaries of the early and late de Sitter epochs, i.e. $a_i$ and $a_f$, are fixed by extending the slopes at the transition points of the Universe where $q=0$, and determining their intersection with the Hubble radius curve \cite{doi:10.1142/S0218271813420017}. Note that the tangent lines drawn at transition points, in $\left(\ln(a),\ln\left(d_{\mathcal{H}}\right)\right)$ coordinate system are always $45$\textdegree\, straight lines, irrespective of the cosmological model\footnote{Slope of a tangent line drawn at a point $a_{\star}$ in $\left(\ln(a),\ln\left(d_{\mathcal{H}}\right)\right)$ coordinate system can be represented in terms of $q$ as,
	\begin{equation}\label{slpe}
		m_{a_\star}= \frac{d \left[\ln\left(d_{\mathcal{H}}\right)\right]}{d \left[\ln(a)\right]}\;\;\Bigg|_{a=a_{\star}}=1+q\,\Big|_{a=a_{\star}}
	\end{equation}
	Since $q=0$ at transition points, we obtain $m_{a_e}=m_{a_t}=1$. Which means, the tangent line drawn at transition points (where $q=0$), in $\left(\ln(a),\ln\left(d_{\mathcal{H}}\right)\right)$ plane always have slope equal to $45$\textdegree.
}.

A perturbation mode with wavenumber `$k$' and wavelength `$\lambda=a/k$' crosses the Hubble radius when it satisfies, $\lambda=d_{H}$, equivalently $k=aH.$ We now have to obtain the number of modes inside the horizon which satisfies this conditions. The number of modes residing inside the comoving volume, $V_{c}=4\pi/[3(aH)^3]$ having wavenumber in the interval $(k,k+dk)$ is,  $dN=V_{c}d^3k/(2\pi)^3.$ Hence, the total number of modes which satisfies the above condition to cross the Hubble radius in the interval $(a_{I}<a<a_{F})$ becomes \cite{doi:10.1142/S0218271813420017},
\begin{equation}\label{N}
	N(a_{I},a_{F})=\left|\int_{a_{I}}^{a_{F}} \frac{V_{c}k^2}{2\pi^2} \frac{dk}{da} da \right|=\frac{2}{3\pi}\left|\ln\left[\frac{a_{F} \mathcal{H}_{F}}{a_{I} \mathcal{H}_{I}}\right]\right|
\end{equation}
In the above relation, an absolute sign is introduced to ensure that $N\geq0$ at all times\footnote{Absolute sign is needed particularly while integrating between the interval $a_1<a<a_2$, during a decelerated expansion phase (radiation-matter era). This is because one gets $a_1H_1<a_2H_2$ during that interval, causing result of the integral to be negative.}. Then, by using Eqns. (\ref{q0}), (\ref{ae}) \& (\ref{late}) in the above expression, one can determine the number of modes that enter the Hubble radius when universe is in between the two de-Sitter expansion periods $(N_c=N(a_{e},a_{t})$) as,
\begin{equation}\label{Nc}
	N_{c}=\frac{2}{3\pi}\left|\ln\left\{\left[\frac{\sqrt[3]{1-\Omega^0_{\Lambda}}\sqrt[6]{\Omega^0_{\Lambda}}}{\sqrt[3]{2}}\right]\sqrt[4]{27\,\Omega^{0}_{\zeta}\sqrt[\gamma]{\Omega^0_{\gamma}}}\right\}\right|
\end{equation}
This shows that the total number of modes re-entering the Hubble sphere between the two de-Sitter epochs, not only depends on the value of all three critical densities ($\Omega^0_{\Lambda}$, $\Omega^0_{\zeta}$, $\Omega^0_{\gamma}$), but on the swiftness parameter ($\gamma$) as well. Imposing $N_c=4\pi$ in Eqn. (\ref{Nc}) we get,
\begin{equation}
	\Omega^0_{\zeta}=\frac{e^{-24\pi^2}}{27\,\sqrt[\gamma]{\Omega^0_{\gamma}}}\left\{\frac{4}{\Omega^0_{\Lambda} \left(1-\Omega^0_{\Lambda}\right)^{2}}\right\}^{2/3}.
\end{equation}
 Considering the observed value of model parameters from Planck analysis \cite{aghanim2020planck}, stated in earlier the sections, we get; $\Omega^0_{\zeta}=7.41 \times 10^{-108}$ which implies $ H_I = 6.10 \times 10^{13}$ GeV.

\section{Scalar field description of the early inflationary phase}\label{3}

In this section we determine the nature of the inflationary expansion predicted by the present model.  This can be determined by studying the evolution of the slowroll parameter, which characterizes the inflation. For finding the evolution of the slowroll parameter, we formulate an equivalent scalar field description of the inflationary phase in the present model. In the presence of a scalar field `$\Phi$' with energy density $\rho_{\Phi}=\dot{\Phi}^2/2 +V(\Phi)$, and pressure $p_{\Phi}=\dot{\Phi}^2/2 -V(\Phi)$, the FLRW equations become, 
\begin{align}
	3H^2=&\,\dot{\Phi}^2/2 +V(\Phi) \label{F1phi}\\
	2\dot{H}+3H^2=&-\dot{\Phi}^2/2 + V(\Phi)\label{F2Phi}
\end{align}
Defining the number of e-folds during the inflationary epoch as, $N=\ln(a/a_i)$ (where, $a_i$ is the scale factor at the beginning of inflation), and using Eqns. (\ref{hi}), (\ref{He}) \& (\ref{ae}), we can then express the Hubble parameter in the early Universe as, 
\begin{equation}\label{HeN}
	\mathcal{H}=\frac{\mathcal{H}_I}{\sqrt{1+e^{\,4\left(N-N_t\right)}}} \quad \textbf{;}\quad \text{with,\;\; } N_t=\ln(a_e/a_i)
\end{equation}
Note that $N_t$ represents the total number of e-foldings from the beginning, to the end of inflation. Subsequently, by combining the Friedmann equations (\ref{F1phi}) \& (\ref{F2Phi}), one obtains the relations,
\begin{align}
	\frac{\Phi}{H_0}=&\,\tilde{\Phi}=\,\pm\sqrt{2}\int_{0}^{\,N} \sqrt{\frac{-\mathcal{H}^{\prime}}{\mathcal{H}}}\; d\tilde{N}\\
	\frac{V(\Phi)}{H_{0}^2}=&\, \tilde{V}=\, 3\mathcal{H}^2 \left[1+ \frac{\mathcal{H}^{\prime}}{3\mathcal{H}}\right] \label{V}
\end{align}
Here, an overhead `\textit{prime}' denotes derivative with respect to e-folding number. Solving these equations by using Eqn. (\ref{HeN}) we get,
\begin{align}
	\tilde{\Phi}&=\,\arctanh \left[\frac{\sqrt{e^{\,4N_t}+1}-\sqrt{e^{\,4 \Delta_N}+1}}{\sqrt{\left(e^{\,4 \Delta_N}+1\right)\left(e^{\,4N_t}+1\right)}-1}\right] \label{phi}\\	\tilde{V}&=\,\mathcal{H}^2_{I}\;\left[\frac{e^{\,4 \Delta_N} \left[3 e^{\,4  \Delta_N}+1\right]}{\left[e^{\,4 \Delta_N}+1\right]^{\,2}}\right] \;\;\textbf{;}\;\; \Delta_N= N_t-N \label{v}
\end{align}
These equations predict a scalar field associated with a Hilltop potential which rolls down from a maximum value of, $\tilde{V}=3\mathcal{H}_{I}^2$ at $N \to 0$ to $\tilde{V}=\mathcal{H}_{I}^2$ at $N \to N_t$. Hence, the value of potential at the beginning of inflation is thrice its value at the end of inflation. Note that this potential, when represented in terms of scale factor, has a similar form as the one in Running Vacuum model \cite{Basilakos:2019zsf}.

\subsection*{Evolution of potential and the slowroll parameters} 
Evolution of slowroll parameters are defined as \cite{PhysRevD.23.347},
\begin{equation}\label{SR}
	\varepsilon_1=\frac{1}{2}\left[\tilde{V}^{\prime}/\tilde{V}\right]^2 \quad \quad \textbf{;}\quad \quad \varepsilon_2= \tilde{V}^{\prime \prime}/\tilde{V}
\end{equation}
Here, `\textit{prime}' denotes derivative with respect to $\Phi$. Using Eqns. (\ref{HeN}) and (\ref{phi}), we can express the potential in terms of scalar field ($\tilde{\Phi}$) and then determine the evolution of slowroll parameters in this model as,
\begin{align}
	&\tilde{V}=\mathcal{H}^2_{I}\left[3-2 \tanh ^2(\hat{\Phi})\right] \left[1-\tanh ^2(\hat{\Phi})\right]\\
	&\varepsilon_1=\frac{\left[17 \tanh (\hat{\Phi})+\sinh (3\hat{\Phi})\; \text{sech}(\hat{\Phi})\right]^2}{2 \left[\cosh (2 \hat{\Phi})+5\right]^2}\\
	&\varepsilon_2=4+\frac{76}{\cosh (2\hat{\Phi})+5}-20\; \text{sech}^2(\hat{\Phi})\\
	&\text{with,} \quad\hat{\Phi}=\tilde{\Phi} + \arctanh\left[\frac{1}{\sqrt{1+e^{4N_t}}}\right]
\end{align}

\begin{figure}
	\centering
	\includegraphics[width=\columnwidth]{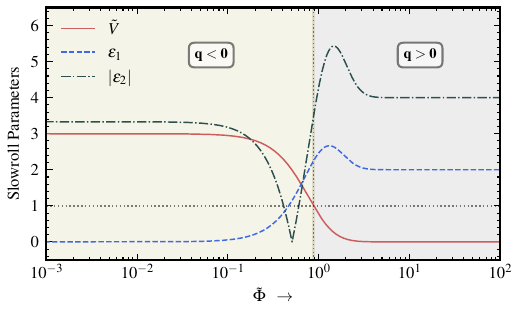}
	\caption{Evolution of scalar potential and slowroll parameters with scalar field, considering $N_{t}=60$.}
	\label{SLRP}
\end{figure}

From the evolution of slowroll parameters as seen in Fig. (\ref{SLRP}), it is evident that, when Universe is well inside the inflationary phase, the value of first slowroll parameter is approximately zero ($|\varepsilon_1|\ll1$), where as the value of second slowroll parameter is well above one ($|\varepsilon_2|\approx3.33$). Consequently, the present model does not imply a conventional slowroll inflation which requires $|\varepsilon_1|\ll1$ and $|\varepsilon_2|\ll1$. Nevertheless, these values are consistent with constant-roll/ultra-slowroll inflation, in which case $\varepsilon<1$ while $|\varepsilon_2|>1$ \cite{PhysRevD.72.023515}. Similar inflationary models have been widely theorized in cosmology \cite{Choudhury:2024one,Motohashi:2014ppa,Dimopoulos:2017ged}, and are considered to be primary seeds to primordial black hole formation, which themselves are argued to be viable dark matter candidates \cite{PhysRevD.81.104019,Montero-Camacho_2019,Ballesteros:2024zdp}. Notably, near to the end of inflation, both parameters eventually becomes less than one for a brief period, indicating a near-slowroll period.

\subsection*{Stability analysis of early de-Sitter solution and the possibility of \enquote{\textit{graceful exit}\,}}

The validity of the present model depends on whether it can predict a smooth transition from the inflationary phase onto the radiation-dominated era. For this, the initial quasi-de-Sitter solution must be unstable state of the system. To investigate this possibility, one considers small perturbations from the initial de Sitter solution, and analyze the evolution of the perturbation modes \cite{Brevik:2017msy}. For a viable inflationary solution, the perturbations in the initial de-Sitter epoch must grow so that it evolves and makes a transition in to the radiation dominated epoch. The perturbed Hubble parameter is taken as \cite{Brevik:2017msy},
\begin{equation}\label{perturb}
	\mathcal{H}=\mathcal{H}_I\left[1 + \Psi(\tau)\right] 
\end{equation}
Here, $\tau=H_{0}t$, is the cosmic time in dimensionless unit and $\Psi(\tau)= \text{exp}\left(\delta \tau\right)$ represents a perturbation from the initial de-Sitter state $\mathcal{H}_I$. It is therefore clear that the deviations from early de-Sitter solution grows only when $\delta>0$. Hence, only the solution that has $\delta>0$ will lead to an unstable de-Sitter state and predict a graceful exit from inflation. Subsequently, we can determine the evolution of scale factor of the Universe as\footnote{To determine the scale factor, we integrate the dimensionless Hubble parameter subject to the initial condition $\tau=0$ at $a=a_i$.},
\begin{equation}
    \quad a =a_i \,e^{\mathcal{H_I}\eta} \quad \quad 
\textbf{;} \quad \quad  \eta=\int_{0}^{\tau} \left[1+\Psi(\tau)\right]\,d\tau \label{scale}
\end{equation}
Combining equations (\ref{ODE}), (\ref{perturb}) and (\ref{scale}) we get,
\begin{multline}\label{perturb2}
	2\mathcal{H}_{I}\delta e^{\delta\tau}=\left[\mathcal{H}_{I}^2\left(1+2e^{\delta \tau}\right)-\Omega^0_{\Lambda}\right]\cdot\\\left\{3\Omega^0_{\zeta}\mathcal{H}_{I}^2\left(1+2e^{\delta \tau}\right) - \left[3+\frac{1}{1+\Omega^0_{\gamma}\,\left(a_i\,e^{\mathcal{H}_{I}\eta}\right)^{\gamma}}\right]\right\}
\end{multline}
Note that, since $|\Psi(\tau)|\ll1$, we have considered only the first order deviations from solution, and have hence used the approximate relation $(1+e^{\delta \tau})^{2} \approx 1+2e^{\delta \tau}$. Also, with $\Psi(\tau)= \text{exp}\left(\delta \tau\right)$ we get, $\eta=\tau+\left[\text{exp}(\delta \tau)-1\right]/\delta$.  

\begin{figure*}
	\centering
	\begin{subfigure}{\columnwidth}
		\centering
		\includegraphics[width=\columnwidth]{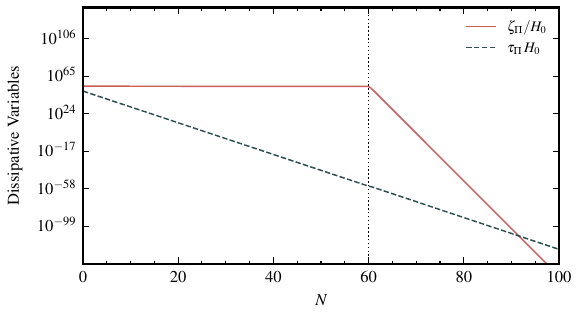}
		\caption{Dissipative variables.}
		\label{DVS}
	\end{subfigure}
	\begin{subfigure}{\columnwidth}
		\centering
		\includegraphics[width=\columnwidth]{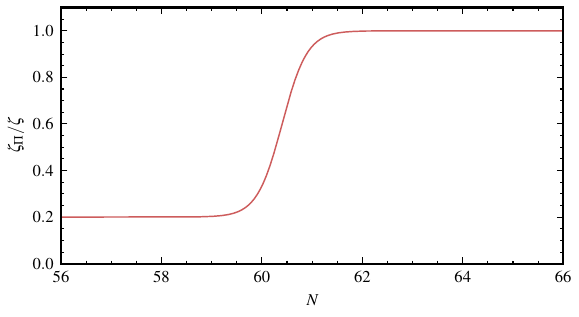}
		\caption{Ratio of viscous coefficients.}
		\label{zetaratio}
	\end{subfigure}
	\caption{Figure on the left shows evolution of viscous coefficient and relaxation time with number of e-folds, and the figure on the right depicts the evolution of ratio of viscous coefficients with number of e-folds. For plotting both these figures we have considered the prior values; $\Omega^0_{\zeta}=10^{-108}$, $\Omega^0_{\gamma}=10^{4}$, $\gamma=1$ and $N_t=60$.}
	\label{DV}
\end{figure*}

To analyze the stability of the early de-Sitter solution in the present model, we consider the limit $\tau \to 0$ in Eqn. (\ref{perturb2}). Simplifying the obtained expression\footnote{Note that, as $\tau\to 0$, we get, $\eta\to0$ and $\Omega^0_{\gamma}\,\left(a_i\,e^{\mathcal{H}_{I}\eta}\right)^{\gamma}\ll1$.}, one can obtain the value of`\,$\delta$\,' as,
\begin{equation}
	\delta= \frac{\left(3\mathcal{H}^2_{I}-\Omega^0_{\Lambda}\right)\left(9\Omega^0_{\zeta}\mathcal{H}^2_{I}-4\right)}{2\mathcal{H}^2_{I}} \; \textbf{;}\;
	\begin{cases}
		\text{\;Since,}& \eta \to 0\\
		\text{\;as}&\tau \to 0
	\end{cases}\notag
\end{equation}
Then, by using Eqn. (\ref{hi}) we finally arrive at,
\begin{equation}\label{delta}
	\delta = \left(4-\Omega^0_{\Lambda}\Omega^0_{\zeta}\right)\sqrt{\frac{12}{\Omega^0_{\zeta}}}
\end{equation}

Analyzing (\ref{delta}) using the prior value of model parameters, we learn that $\delta>0$. Hence, the early de Sitter solution is indeed an unstable state of the system, and consequently, this model offers a graceful exit from early inflation to radiation dominated epoch. We also find that the value of $\delta$ is significantly large, and is of the order of $\mathcal{H}_{I}$ itself, i.e., $\delta\approxeq12\,\mathcal{H}_{I}$. Which means, perturbations in the early epoch grows exponentially at a rate that is approximately twelve times the Hubble rate.

\section{Evolution equation for bulk viscosity in the extreme causal limit} \label{isp}

Up to this point, the bulk viscous pressure in this model was characterized by standard Eckart relation, $\Pi=-\zeta \nabla_\mu u^\mu$. Even though this first-order relation is the most straightforward  relativistic extension of the Navier–Stokes equation, it is well established that this formulation violates causality \cite{ISRAEL1979341}. Nonetheless, since the present framework exhibits rich dynamical behavior that unifies all major epochs of cosmic evolution, it is important to examine whether a causal viscous evolution can reproduce the same background dynamics without introducing additional free parameters. This section is devoted to formulating such a causal extension for the bulk viscosity.

Direct way to do this is to construct a Israel–Stewart type causal evolution equation for the bulk viscosity that yields the same background dynamics as that obtained from the Eckart relation. In this formulation, the bulk viscosity coefficient $\zeta$ appearing in first order relation is reinterpreted as an effective viscous coefficient emerging from an underlying non-linear Israel–Stewart differential equation with transport variables $\zeta_\Pi$ and $\tau_\Pi$, both of which can also depend on the viscous pressure. Using the causality constraint proposed in \cite{PhysRevLett.122.221602}, together with the Eckart relation, we then derive a unique and exact relation connecting these re-defined transport variables to the effective coefficient $\zeta$. For physical consistency, these model parameters must satisfy the non-negativity conditions, $\zeta_\Pi \geq 0$ and $\tau_\Pi \geq 0$ throughout the cosmic evolution. In the following analysis, since we focus on the inflation–radiation transition epoch where viscous effects are the most significant, we adopt the approximations $\omega \simeq 1/3$ and $\Lambda/H^2 \ll 1$, which are valid in this regime.

Using the relation, $\Pi=-\zeta_0\rho^2/H_p^2$, its time derivative, along with Eqns. (\ref{F1}) and (\ref{F2}), we can construct a first order differential equation in viscous pressure as,
\begin{equation}
	\tau_\Pi \dot{\Pi} + \Pi = -\frac{3\zeta_0 H}{H_p^2}\sqrt{\frac{\rho^{\,3}}{3}}\left[ 1-2\tau_{\Pi}\sqrt{\frac{\rho}{3}}\left(\frac{4}{3}+\frac{\Pi}{\rho}\right)\right]
\end{equation}
Here, $\tau_\Pi$ is a dynamical parameter associated with relaxation time of the fluid. Comparing this expression with the Israel-Stewart equation, $\tau_\Pi \dot{\Pi} + \Pi = -3\zeta_\Pi H$, one can identify a new underlying viscous coefficient,
\begin{equation}\label{zetaPI}
	\zeta_\Pi=\frac{\zeta_0}{H_p^2}\sqrt{\frac{\rho^{\,3}}{3}}\left[ 1-2\tau_{\Pi}\sqrt{\frac{\rho}{3}}\left(\frac{4}{3}+\frac{\Pi}{\rho}\right)\right]
\end{equation}
Note that, $\tau_\Pi$ is still an unknown parameter in this case. However, we find that for satisfying causality, at least in the extreme limit, $\zeta_\Pi$ and $\tau_\Pi$ must be such that\footnote{Here, we have considered the causality to be satisfied in the extreme limit by treating the expression (1) in \cite{PhysRevLett.122.221602} as an equality.},
\begin{equation}\label{caus}
	\zeta_{\Pi}= \frac{2}{9}\left(4\rho+3\Pi\right)\tau_{\Pi}
\end{equation}
Therefore, in its extreme limit, the causality constraint imposed on the system uniquely specifies the relationship connecting the relaxation time of the fluid and the bulk viscous coefficient, without adding any new parameters in the model. The evolution of underlying bulk viscous variables are then obtained as,
\begin{align}
	\zeta_{\Pi}/H_0=&\;\Omega^0_{\zeta}\mathcal{H}^3\left[ 1+3\Omega^0_{\zeta}\mathcal{H}^2\right]^{-1}\\
	H_0\tau_{\Pi}=&\; \frac{3\Omega^0_{\zeta}\mathcal{H}}{2\left[3\Omega^0_{\zeta}\mathcal{H}^2-4\right]\left[\, 1+3\Omega^0_{\zeta}\mathcal{H}^2\right]}
\end{align}
Accordingly, we see that if, $|\Pi|/\rho < 4/3$, which is indeed the case, the present model satisfies the thermodynamic constraints $\tau_{\Pi}\geq0$ and $\zeta_{\Pi}\geq 0$ at all times. Hence, the causal evolution equation obtained above is indeed a viable transport equation for bulk viscous pressure.

The evolution of dissipative variables predicted by the above equations, and the ratio of viscous coefficients, are provided in Fig. (\ref{DV}). Accordingly, we see that during inflation, (a) the value of both, the viscous coefficient, and the relaxation time is significantly large compared to their values in the radiation-dominated epoch, (b) $\zeta$ and $\zeta_{\Pi}$ differ considerably from one another. However, as the universe transits from inflation to radiation-dominated era, the values of $\tau_{\Pi}$ and $\zeta$ decay extremely rapidly, while $\zeta_\Pi$ approaches $\zeta$. Hence, as the universe transitions from inflation to radiation, the transport equation for bulk viscous pressure relaxes from Israel-Stewart relation to Eckart relation. Moreover, since the causal transport equation signifies viscoelastic behavior for the fluid, while the first order relation represents pseudoplastic nature \cite{PhysRevD.109.096040}, the transition from inflation to radiation dominated epoch directly implies a transition in viscous nature of the cosmic fluid, from viscoelastic to pseudoplastic.

\section{Summary}\label{5}

In this article, we extended the standard $\Lambda$CDM model by incorporating bulk viscosity emerging from isentropic particle creation mechanism in the early universe, and derived a novel analytical model for Hubble parameter that provides a unified description of the entire cosmic evolution from the initial inflationary phase to late accelerating epoch, including the intermediate radiation and matter-dominated eras. This was achieved by considering the cosmic medium as a bulk viscous cosmic fluid (vCF) governed by an evolving adiabatic equation of state, and a viscous coefficient that is directly proportional to both, the energy density of vCF, and the expansion velocity. These two basic, physically motivated assumptions allows for a self-consistent evolution of the Hubble parameter across all epochs. Using the obtained analytical solution, we determined the finite, physically relevant boundary of both de Sitter epochs, and the expression for the number of perturbation modes crossing the Hubble volume between those two de Sitter phases (called the Cosmic Mode Index or CosMIn, by Padmanabhan). Furthermore, by using the postulate that CosMIn is an epoch-invariant quantity with a value $N_c=4\pi$ \cite{doi:10.1142/S0218271813420017}, we proposed a novel method to extract the value of viscous coefficient (and by extension the inflation energy scale) in the present model. This formulation naturally establishes a direct relationship between the inflationary energy scale and the other free parameters of the theory. Considering the parameter values reported by the Planck Collaboration \cite{aghanim2020planck}, we obtain an inflationary Hubble parameter of $H_I \approx 6.10 \times 10^{13}$ GeV,
which lies well within the range that is consistent with current observational constraints.

Exact nature of the inflationary expansion is then investigated by recasting the early time dynamics in the model in terms of a single scalar-field and determining the evolution of slow-roll parameters. Obtained results suggests an ultra-slowroll Hilltop inflationary expansion with a near-slowroll period towards the end of inflation, and a graceful exit to radiation dominated epoch. There are recent studies which suggest that ultra-slow-roll inflation may play a crucial role in seeding primordial black holes \cite{PhysRevLett.133.121403,GU2025101744}, thereby enhancing the model's cosmological relevance and motivating further research. Finally, by invoking the causality constraint in the extreme limit during the viscous driven early universe, we determined an equivalent causal transport equation for bulk viscosity that predicts identical background evolution for the Universe, compared to the first order Navier-Stokes relation. Further analysis revealed that, during the transition from the inflationary phase to the radiation-dominated epoch, the characteristic nature of the vCF undergoes a qualitative change, evolving from viscoelastic to pseudoplastic behavior. However, it must be noted that, even though both relations predict exactly identical background evolution for bulk viscosity, they can differ dramatically under perturbations, and hence, for detailed perturbative studies in the future, the causal evolution equation is recommended to ensure well behaved solutions.

\section*{Acknowledgments}
Analytical calculations were performed using Wolfram Mathematica, Version 14.2, under a 15-day trial license provided by Wolfram Research \cite{Wolfram|AlphaNotebookEdition}. Vishnu A Pai is thankful to Cochin University of Science and Technology for providing Senior Research Fellowship. This research was supported in part by the International Centre for Theoretical Sciences (ICTS) through the program:- \textit{3rd IAGRG School on Gravitation and Cosmology}. (code: ICTS/iagrg2024/10).

%\bibliography{ref.bib}

%

\end{document}